\documentclass[12pt,graphicx,amssymb,amsmath,amsthm,amsfonts]{iopart}

\usepackage[usenames]{color}
\usepackage{cprotect}
\usepackage{wasysym}
\usepackage{hyperref}
\usepackage{cite}
\usepackage{amsfonts}
\usepackage{tensor,multirow}

\newcommand{\floor}[1]{\lfloor#1\rfloor}

\begin{document}


\title{General spherical harmonic bra-ket overlap integrals of trigonometric functions}

\author{Giuseppe Lingetti\footnote{giuseppe.lingetti@uniroma1.it}}
\address{Dipartimento di Fisica, ``Sapienza" Università di Roma \& Sezione INFN Roma1, Piazzale Aldo Moro 5, 00185, Roma, Italy}
\author{Paolo Pani}
\address{Dipartimento di Fisica, ``Sapienza" Università di Roma \& Sezione INFN Roma1, Piazzale Aldo Moro 5, 00185, Roma, Italy}


\begin{abstract}
Closed formulas in terms of double sums of Clebsch-Gordan coefficients are computed for the evaluation of bra-ket spherical harmonic overlap integrals of a wide class of trigonometric functions. These analytical expressions can find useful application in problems involving non-separable wave equations, e.g. general-relativistic perturbation theory, electromagnetism, quantum theory, etc, wherein the overlap integrals arise from the coupling among different angular modes. We provide some examples related to linear perturbations of spinning black holes in General Relativity and modified gravity, in which the analytical formulas for the overlap integrals are particularly useful to compute the black-hole spectrum.
\end{abstract}



\section{Introduction}

The spherical harmonic decomposition is a widely used tool in perturbation theory. Applications are countless and include any linear theory (e.g., electromagnetism~\cite{Jackson}, quantum mechanics~\cite{Sakurai}, Newtonian gravity~\cite{PoissonWill}), and any perturbative scheme of nonlinear theory (e.g., black-hole perturbation theory within and beyond General Relativity~(GR)~\cite{Pani:2013pma}).
The spherical harmonic decomposition finds its main advantage for spherically symmetric systems in order to separate the radial and angular part of the perturbation equations, but it can also be applied to less symmetric configurations. While the angular modes decouple from each other in spherical symmetry --~leading to lower-dimensional differential equations~-- they typically mix when spherical symmetry is broken. This mode-mixing gives rise to bra-ket integrals that must be explicitly evaluated~\cite{Pani:2013pma}. 

A relevant class of these braket integrals involves trigonometric functions:
\begin{equation}\label{bra_ket1}
    \left\langle l_1, m_1 \right| e^{i k \phi} \cos(n \theta) \left| l_2 , m_2\right\rangle=
\end{equation} 
\begin{equation*}
    ~~~~~~~~~~~~~~~~~~~~~~~~~~~~=\int\limits_0^{2 \pi}d\phi\int\limits_0^\pi d\theta \sin\theta Y^*_{l_1 ,m_1}(\theta,\phi) e^{i k \phi} \cos(n \theta)Y_{l_2 ,m_2}(\theta,\phi)\,,
\end{equation*}
\begin{equation}\label{bra_ket2}
    \left\langle l_1, m_1 \right| e^{i k \phi} \sin(n \theta) \left| l_2 , m_2\right\rangle=\nonumber
\end{equation} 
\begin{equation*}
    ~~~~~~~~~~~~~~~~~~~~~~~~~~~~=\int\limits_0^{2 \pi}d\phi\int\limits_0^\pi d\theta \sin\theta Y^*_{l_1 ,m_1}(\theta,\phi) e^{i k \phi} \sin(n \theta)Y_{l_2 ,m_2}(\theta,\phi)\,,
\end{equation*}
Due to the fact that $e^{i k \phi} \cos(n \theta)$ and $e^{i k \phi} \sin(n \theta)$ form a complete Fourier basis for $(\theta,\phi)\in\left[0,\pi\right]\times\left[0,2\pi\right)$, from the general solutions of Eqs.~(\ref{bra_ket1}) and~(\ref{bra_ket2}) in principle one can directly compute the bra-kets of any trigonometric function regular in such interval \footnote{In principle, through a Fourier expansion, one could also apply the general solutions to angular functions which are not strictly trigonometric. However, the feasibility of such approach clearly depends on how easily the Fourier expansions can be computed for the problem at hand.}. Overlap integrals of this type naturally arise from the spherical harmonic decomposition of wave equations, such as those coming from GR perturbation theory~\cite{Pani:2013pma}. Important examples are bosonic perturbations of spinning black holes~\cite{dolan_time_domain,Baumann_2019}, or the linear dynamics of ``dirty'' black holes in GR~\cite{Dima_2020} and in extended theories of gravity~\cite{scalar-tensor,Cardoso:2013opa,Lingetti:2022psy}. In the easiest cases having the analytical results helps in understanding the type of couplings involved: for example from the result of $\left\langle l_1, m_1 \right|\cos^2 \theta \left| l_2 , m_2\right\rangle$, one can easily deduce that scalar perturbations of a Kerr black hole have couplings between $l$ and $l\pm2$ modes~\cite{dolan_time_domain}. In the most cumbersome cases, instead, the number of different couplings can be much higher and/or show involved expressions~\cite{Pani:2013pma,Baumann_2019}, thus closed-form solutions can greatly speed up computations. 

To the best of our knowledge, in the literature there is no general, analytical, solution to the bra-ket trigonometric integrals~(\ref{bra_ket1}) and~(\ref{bra_ket2}). In this short note we provide a closed-form general solution by exploiting the properties of ultraspherical polynomials and their relationships with trigonometric functions and spherical harmonics. 

Before deriving the full expression for the bra-ket trigonometric integrals~(\ref{bra_ket1}) and~(\ref{bra_ket2}) in Sec.~\ref{sec:derivation}, in Sec.~\ref{sec:examples} we give some specific examples wherein these integrals naturally emerge.
The appendices are devoted to an alternative derivation of the same integrals.


\section{Some examples}\label{sec:examples}

Two notable examples in which the integrals~(\ref{bra_ket1}) and~(\ref{bra_ket2}) emerge are massive bosonic perturbations of Kerr black holes~\cite{Brito:2015oca} and the linearized dynamics of Kerr black holes in scalar-tensor theories interacting with matter~\cite{Lingetti:2022psy}. We briefly discuss these examples in the next subsections.

\subsection{Massive spin-1 perturbations of a Kerr black hole}

Massive bosonic perturbations of a spinning  black hole play an important role for the stability analysis of the Kerr metric and in the context of superradiant instabilities triggered by ultralight bosonic fields~\cite{Arvanitaki, kodama, yoshino, BH_bomb, Perturbs_pani, Baryakhtar,brito_1,brito_2} While the spin-0 case is directly separable in terms of spheroidal harmonics, only a few years ago was the spin-1 case shown to be separable~\cite{frolov_proca}.
Here, to highlight the use the overlap integrals we will not exploit the separability, also because the latter can be used only in the frequency domain (see Ref.~\cite{Dolan:2012yt} for a similar discussion in the scalar case).

Massive vector perturbations of the Kerr metric are described by the Proca field equations in a Ricci-flat\footnote{Note that the described method is generic and does not rely on the special symmetries of the Kerr metric, so it can be applied also in different situations.} spacetime background,
\begin{equation}\label{proca}
    \Box A_\nu - \mu^2 A_\nu=0~,\qquad \nabla^\nu A_\nu=0\,,
\end{equation}
where $A_\nu$ is the vector perturbation, $\mu$ is proportional to its mass.
By following \cite{Baumann_2019}, we can perform a spherical harmonic decomposition of field equations (\ref{proca}), thus getting an infinite cascade of coupled radial equations. The coupling terms are bra-ket integrals of rational functions $X$ arising from the metric and its Christoffel symbols, functions whose dependence on $\theta$ is through trigonometric functions $\cos\theta$ and $\sin\theta$ while they depend on $\phi$ through some phase functions. Using Boyer-Lindquist coordinates $(t, r, \theta, \phi$), we can generically label such functions as $X(r, \theta, e^{i\phi}; a, M)$, where $a<M$ is the black hole spin parameter, $M$ is the black hole mass. For $a=0$ these functions are linear combinations of trigonometric functions in $\theta$, while for $a\neq 0$ they are cumbersome rational functions involving $\cos\theta$ and $\sin\theta$. Brute force computation of all the integrals $\left\langle l_1, m_1 \right|X(r, \theta, e^{i\phi}; a, M)\left| l_2 , m_2\right\rangle$ is very time consuming and must be performed by fixing the parameters of the system, which implies that any change in the parameters requires recomputing those integrals from scratch. Furthermore, the number of relevant overlapping integrals can be very high, especially for highly spinning black holes. Indeed, if we truncate our spherical harmonic expansion such that $l_1, l_2\leq L$ for some value $L>0$, the number of overlap integrals to be computed will be proportional to $L^2 (L+2)^2$\footnote{If we just consider $\left|l,m \right\rangle$ for $0\leq l\leq L$, we have $\sum\limits_{l=0}^L (2 l+1)=L(L+2)$ possible spherical harmonics. Each integral involves two spherical harmonics, thus the number of integrals is proportional to $L^2(L+2)^2$}.

A way to bypass such problems is to Fourier-expand $X(r, \theta, e^{i\phi}; a, M)$, in which case the problem is reduced to computing integrals (\ref{bra_ket1}) and (\ref{bra_ket2}):
\begin{equation}\label{Fourier-expand}
\left\langle l_1, m_1 \right| X(r, \theta, e^{i\phi}; a, M) \left| l_2 , m_2 \right\rangle=
\end{equation}
\begin{equation*}
~~~~~~~~~~~~~~~~~~=\sum\limits_{n=0 , k}^\infty \tilde{X}_{n,k}(M,a; r)\left\langle l_1, m_1 \right| e^{i k \phi} \cos(n \theta) \left| l_2 , m_2\right\rangle+
\end{equation*}
\begin{equation*}
~~~~~~~~~~~~~~~~~~+\sum\limits_{n=1 , k}^\infty \tilde{X}_{-n,k}(M,a; r)\left\langle l_1, m_1 \right| e^{i k \phi} \sin(n \theta) \left| l_2 , m_2 \right\rangle
\end{equation*}
In order to compute the Fourier coefficients $\tilde{X}_{n,k}$ without carrying any integration, one can resort to a Taylor expansion of $X(r, \theta, e^{i\phi}; a, M)$ in the black-hole spin parameter at arbitrarily high order\footnote{For accurate results, one needs at least order equal to $2 L$ in the spin expansion.}. Due to how the spin parameter is coupled to the trigonometric functions, once the integrals (\ref{bra_ket1}) and (\ref{bra_ket2}) are computed analytically it is possible to derive analytical expressions for $\tilde{X}_{n,k}$ at arbitrarily high order in the black-hole spin. 

At least in spirit, this approach can be extended also to massive spin-2 perturbations, whose case shows even more cumbersome $X(r, \theta, e^{i\phi}; a, M)$ functions and consequently the computation time for their brute force evaluation would be much higher~\cite{LingettiThesis}. Indeed, at the moment the massive spin-2 case is not separable and hence one has to resort to more sophisticated numerical approaches~\cite{LingettiThesis}.

\subsection{Perturbations of a Kerr black hole in scalar-tensor theories with matter interactions}

A wide class of scalar-tensor theories with one scalar field can be described by the following general action in the Einstein frame \cite{Fujii:2003pa,Dima:2020rzg,Lingetti:2022psy}:
\begin{equation}
    S=\int d^4x \sqrt{-g}\left[\frac{R}{16\pi}-\frac{1}{2}g_{\mu\nu}\partial^\mu \Phi \partial^\nu \Phi-\frac{V(\Phi)}{16 \pi}\right]+ S(\psi_m, \mathcal{A}(\Phi)^2g_{\mu\nu})\,.
\end{equation}
where $g_{\mu\nu}$ is the metric tensor, $R$ is its Ricci scalar curvature, $\Phi$ a scalar field and $V(\Phi)$ some self-interaction for $\Phi$, while the last term describes the presence of some matter fields $\psi_m$. The function $\mathcal{A}(\Phi)$ models the non-minimal coupling of the scalar field to matter: in the Einstein frame the scalar is minimally coupled to the gravity sector, while matter is coupled to the effective metric \textbf{$\mathcal{A}(\Phi)^2g_{\mu\nu}$}.
If we expand the scalar field equations for $\varphi\equiv\Phi-\Phi^{(0)}\ll1$, where $\Phi_0$ is a GR solution, on a Kerr background we get a Klein-Gordon equation with an effective mass squared term \cite{Cardoso:2013opa, Cardoso:2013fwa, Lingetti:2022psy}
\begin{equation}
\label{eq:finalKGeq}
    [\Box-\mu_{\rm eff}^2(r, \theta)]\varphi=0\,.
\end{equation}
where $\mu_{\rm eff}^2(r, \theta)$ depends on the specific model for the matter fields.
For example, Ref.~\cite{Lingetti:2022psy} studies the case of accretion disks around a Kerr black hole. In this case it is convenient to parametrize\cite{Dima:2020rzg, Lingetti:2022psy}
\begin{equation}
    \mu_{\rm eff}^2(r,\theta)=\mu_r^2(r)\sin^{2 q}\theta+\mu_0^2(r)
\end{equation} for some positive integer $q$ and radial functions $\mu_r(r)$ and $\mu_0(r)$. The higher the $q$ parameter is, the thinner the accretion disk around the equatorial plane. Such type of functions generate overlap bra-ket integrals which can easily be written as linear combinations of the integrals (\ref{bra_ket1}). More involved models for $\mu_{\rm eff}^2(r, \theta)$ might be treated like the massive spin-1 case, but the feasibility of computing the associated Fourier expansions should be estimated on a case-by-case basis. 
\section{Evaluation of the bra-ket integrals} \label{sec:derivation}

Having presented some relevant examples in which the bra-ket integrals (\ref{bra_ket1}) and (\ref{bra_ket2}) naturally appear, we can now derive their expression in a closed form.

We recall that spherical harmonics are defined as~\cite{rose} 
\begin{equation}\label{spherical_harmonics}
Y_{l,m}(\theta,\phi)=\sqrt{\frac{(2 l+1)(l-m)!}{4\pi(l+m)!}}P_l^m(\cos\theta)e^{i m \phi}\,,
\end{equation}
where $P_l^m$ are the Legendre associated functions, and consequently their complex conjugate is $Y_{l , m}^*=(-1)^m Y_{l,-m}$. 
Moreover, we can express their products in the following way~\cite{rose}:
\begin{equation}\label{harmonics_product}
Y_{l_1 , m_1}Y_{l_2 , m_2}=\sum\limits_{l=|l_1-l_2|}^{l_1+l_2}\sqrt{\frac{(2 l_1+1)(2 l_2+1)}{4\pi(2 l+1)}}\left\langle l_1 , 0 , l_2 , 0 \right| \left. l , 0\right\rangle
\end{equation} 
\begin{equation*}
~~~~~~~~~~~~~~~~~~\times\left\langle l_1 , m_1 , l_2 , m_2 \right| \left. l , m_1+m_2\right\rangle Y_{l , m_1+m_2}\nonumber\,,
\end{equation*}
where $\left\langle l_1 , m_1 , l_2 , m_2 \right| \left. l , m\right\rangle$ are the Clebsch-Gordan coefficients.\\

By exploiting these properties, bra-kets~(\ref{bra_ket1}) and~(\ref{bra_ket2}) can be written as

\begin{equation}\label{cos_braket0}
\left\langle l_1, m_1 \right| e^{i k \phi} \cos(n \theta) \left| l_2 , m_2\right\rangle=
\end{equation}
\begin{equation*}
~~~~~~~~~~~~~~~=\sum\limits_{l=|l_1-l_2|}^{l_1+l_2}\sqrt{\frac{\left( l_1+\frac{1}{2}\right)\left( l_2+\frac{1}{2}\right)(l+k)!}{(l-k)!}}(-1)^{m_1}\left\langle l_1 , 0 , l_2 , 0 \right| \left. l , 0\right\rangle
\end{equation*}
\begin{equation*}
~~~~~~~~~~~~~~~~~~\times\left\langle l_1 , -m_1 , l_2 , m_2 \right| \left. l , -k\right\rangle\int_0^\pi d\theta \sin\theta \cos(n\theta) P_l^{-k}(\cos\theta)\,,
\end{equation*}

\begin{equation}
\label{sin_braket0}\left\langle l_1, m_1 \right| e^{i k \phi} \sin(n \theta) \left| l_2 , m_2\right\rangle=
\end{equation}
\begin{equation*}
~~~~~~~~~~~~~~~=\sum\limits_{l=|l_1-l_2|}^{l_1+l_2}\sqrt{\frac{\left( l_1+\frac{1}{2}\right)\left(l_2+\frac{1}{2}\right)(l+k)!}{(l-k)!}}(-1)^{m_1}\left\langle l_1 , 0 , l_2 , 0 \right| \left. l , 0\right\rangle
\end{equation*}
\begin{equation*}
~~~~~~~~~~~~~~~~~~\times\left\langle l_1 , -m_1 , l_2 , m_2 \right| \left. l , -k\right\rangle \int_0^\pi d\theta \sin\theta \sin(n\theta) P_l^{-k}(\cos\theta)\,.
\end{equation*}
Functions $\cos(n\theta)$ and $\sin(n\theta)$ can be expressed as finite linear combinations of associated Legendre functions:
\begin{equation}\label{cos-sin-legendre0}
\cos(n\theta)=\sum\limits_{l=0}^n a^n_l P_l^0(\cos\theta)\,,\qquad\sin(n\theta)=\sum\limits_{l=0}^n b^n_l P_l^1(\cos\theta)\,.
\end{equation}
From the definition of the associated Legendre functions~\cite{math_meth}, we have $P_l^0(\cos\theta)=P_l(\cos\theta)$ and $P_l^1(\cos\theta)=\frac{d}{d\theta}\left[P_l(\cos\theta)\right]$, where $P_l$ are the Legendre polynomials. Thus, by taking the derivative of the first equation in~(\ref{cos-sin-legendre0}) with respect to $\theta$ and comparing it with the second equation, we get
\begin{equation}\label{cos-sin-coeff}
b^n_l=-\frac{1}{n}a^n_l\,.
\end{equation}
One can therefore focus only on the first equation in~(\ref{cos-sin-legendre0}). Functions $\cos(n\theta)$ are Čebyšëv polynomials $T_n(\cos\theta)=\cos(n\theta)$~\cite{special_polynomials}, hence we can exploit this property for finding the $a^n_l$ coefficients. Functions $T_n(x)$ and $P_l(x)$ are in fact special cases of ultraspherical polynomials $C^\gamma_n(x)$~\cite{special_polynomials}:
\begin{equation}\label{ultraspherical}
T_n(x)=\lim\limits_{\gamma\rightarrow 0}\frac{n+2\gamma}{2\gamma}C^\gamma_n(x)\,,\qquad P_l(x)=C^{1/2}_l(x)\,,
\end{equation}
Different types of ultraspherical polynomials can be related through the following expression~\cite{special_polynomials}, which is a special case of the connection relation for Jacobi polynomials\footnote{Ultraspherical polynomials are a special case of Jacobi polynomials~\cite{special_polynomials}.}:
\begin{equation}\label{connection_jacobi}
C^\gamma_n(x)=\sum\limits_{j=0}^{\floor{n/2}}\frac{(\gamma-\beta)_j\left(\gamma\right)_{n-j}}{j!\left(\beta+1\right)_{n-j}}\left(\frac{\beta+n-2 j}{\beta} \right) C^\beta_{n-2 j}(x)\, ,
\end{equation}
where $\displaystyle\left(x \right)_y=\frac{\Gamma(x+y)}{\Gamma(x)}$ is the Pochhammer's symbol for $x,y\in\mathbb{C}$ and $\Gamma(x)$ is Euler's gamma function. From Eqs.~(\ref{ultraspherical}) and~(\ref{connection_jacobi}) we get the expressions for Čebyšëv polynomials as linear combinations of Legendre polynomials
\begin{equation}\label{chebyshev-legendre}
T_n(x)=-\sum\limits_{j=0}^{\floor{n/2}}\frac{n\Gamma\left(j-1/2\right)\Gamma\left(n-j\right)}{8 j!\Gamma\left(3/2+n-j\right)}(1+2n-4j)P_{n-2 j}(x)\,,
\end{equation}
and consequently, by using Eq.~(\ref{cos-sin-coeff}), we can rewrite Eq.~(\ref{cos-sin-legendre0}) as
\begin{equation}\label{cos-legendre}
\cos(n\theta)=-\sum\limits_{j=0}^{\floor{n/2}}\frac{n\Gamma\left(j-1/2\right)\Gamma\left(n-j\right)}{8 j!\Gamma\left(3/2+n-j\right)}\left(1+2 n-4 j \right)P^0_{n-2j}(\cos\theta)\,,
\end{equation}
\begin{equation}\label{sin-legendre}
\sin(n\theta)=\sum\limits_{j=0}^{\floor{n/2}}\frac{\Gamma\left(j-1/2\right)\Gamma\left(n-j\right)}{8 j!\Gamma\left(3/2+n-j\right)}\left(1+2 n-4 j \right)P^1_{n-2j}(\cos\theta).
\end{equation}
By using these last results, we can write the general parametric expressions for the bra-kets~(\ref{cos_braket0}) and~(\ref{sin_braket0}) in closed form:
\begin{equation}\label{cos_braket}
\left\langle l_1, m_1 \right| e^{i k \phi} \cos(n \theta) \left| l_2 , m_2\right\rangle=~~~~~~~~~~~~~~~~~~~~~~~~~~~~~~~~~~~~~~~~~~~~~~~~~~~~~~~~~~~~~~~~
\end{equation}
\begin{equation*}
    =\sum\limits_{l=|l_1-l_2|}^{l_1+l_2} \sum\limits_{j=0}^{\floor{n/2}}\frac{(-1)^{m_1+1}}{2}\frac{n\Gamma\left(j-1/2\right)\Gamma\left(n-j\right)}{8 j!\Gamma\left(3/2+n-j\right)}\sqrt{\frac{(2 l_1+1)(2 l_2+1)(l+k)!}{(l-k)!}}
\end{equation*}
\begin{equation*}
\times  \left(1+2 n-4 j \right) \left\langle l_1 , -m_1 , l_2 , m_2 \right| \left. l , -k\right\rangle \left\langle l_1 , 0 , l_2 , 0 \right| \left. l , 0\right\rangle I(n-2 j, 0, l,-k)\,,
\end{equation*}
\begin{equation}\label{sin_braket}
\left\langle l_1, m_1 \right| e^{i k \phi} \sin(n \theta) \left| l_2 , m_2\right\rangle=~~~~~~~~~~~~~~~~~~~~~~~~~~~~~~~~~~~~~~~~~~~~~~~~~~~~~~~~~~~~~~~~
\end{equation}
\begin{equation*}
=\sum\limits_{l=|l_1-l_2|}^{l_1+l_2} \sum\limits_{j=0}^{\floor{n/2}}\frac{(-1)^{m_1}}{2}\frac{\Gamma\left(j-1/2\right)\Gamma\left(n-j\right)}{8 j!\Gamma\left(3/2+n-j\right)}\sqrt{\frac{(2 l_1+1)(2 l_2+1)(l+k)!}{(l-k)!}}
\end{equation*}
\begin{equation*}\times \left(1+2 n-4 j \right) \left\langle l_1 , -m_1 , l_2 , m_2 \right| \left. l , -k\right\rangle \left\langle l_1 , 0 , l_2 , 0 \right| \left. l , 0\right\rangle I(n-2 j, 1, l,-k)\,,
\end{equation*}
where we have $\displaystyle I(l,m,l',m')=\int_{-1}^1 dx P_l^m(x) P_{l'}^{m'}(x)$. This last integral can be explicitly performed~\cite{legendre_integral}:
\begin{equation*}
I(l,m,l',m')=\sqrt{\frac{(l+m)!(l'+m')!}{(l-m)!(l'-m')!}}\sum\limits_{j=|l-l'|}^{l+l'}\sqrt{\frac{(j-m-m')!}{(j+m+m')!}}
\end{equation*}
\begin{equation*}
~~~~~~~~~~~~~~~~~~\times\left\langle l , m , l', m' \right| \left. j , m+m'\right\rangle\left\langle l , 0 , l', 0 \right| \left. j , 0\right\rangle I_0(j,m+m')\,,
\end{equation*}
where for $m>0$ we have
\begin{equation*}
I_0(l,m)=\int_{-1}^1 dx P_l^m(x)=\frac{[(-1)^m+(-1)^l]2^{m-2}m\Gamma\left(l/2 \right)\Gamma\left((l+m+1)/2\right)}{\left((l-m)/2 \right)!\Gamma\left((l+3)/2 \right)}.
\end{equation*}
Because of $P_0^0(\cos\theta)=1$, for $l,m=0$ we have $I_0(l,m)=2$, while for $m<0$ we can exploit $\displaystyle I_0(l,m)=(-1)^m\frac{(l+m)!}{(l-m)!}I_0(l,-m)$ from the properties of $P_l^m$. Expression~(\ref{cos_braket}) gives an indeterminate result for $n=0$, therefore, by taking into account that $\cos(0 \theta)=1=P_0^0(\cos\theta)$, for this case we need to use the following expression:
\begin{equation}\label{n0cos_braket}\left\langle l_1, m_1 \right| e^{i k \phi} \left| l_2 , m_2\right\rangle=\sum\limits_{l=|l_1-l_2|}^{l_1+l_2}\frac{(-1)^{m_1}}{2} \sqrt{\frac{(2 l_1+1)(2 l_2+1)(l+k)!}{(l-k)!}}
\end{equation}
\begin{equation*}~~~~~~~~~~~~~~~~~~~~~~~~~~~\times\left\langle l_1 , -m_1 , l_2 , m_2 \right| \left. l , -k\right\rangle \left\langle l_1 , 0 , l_2 , 0 \right| \left. l , 0\right\rangle I_0(l,-k)\,.
\end{equation*}

\section{Conclusions}

We derived the general analytical expressions for the spherical harmonic overlap integrals~(\ref{bra_ket1}) and~(\ref{bra_ket2}) of a wide class of trigonometric functions, from which solutions for all continuous trigonometric functions can be easily derived. We expressed the results in closed form in terms of double sums of Clebsch-Gordan coefficients. 
These results can be useful in the computational study of wave equations for non-separable systems, e.g. in perturbed quantum mechanical problems or in the context of spherical harmonic decompositions within black-hole perturbation theory.


\begin{appendix}
\section{Alternative derivation}\label{referee_alternative}

An anonymous referee, to whom we express our gratitude, suggested the following alternative derivation for integrals (\ref{bra_ket1}) and (\ref{bra_ket2}). Legendre polynomials can be expanded as linear combinations of Čebyšëv polynomials via a Fourier transform, whose coefficients can be found in Example~15.1.2 of \cite{whittaker_watson_1996}:
\begin{equation}\label{Lagrange_expand}
    P_l(\cos\theta)=\sum\limits_{j=0}^{\floor{l/2}} a_{l,j} T_{l-2j}(\cos\theta),\quad 
    a_{l,j}=\frac{2 (2l-2j-1)!!(2j-1)!!}{(1+\delta_{l-2j,0})(2l-2j)!!(2j)!!}\,.
\end{equation}
Thus, by exploiting the properties of the product of Čebyšëv polynomials\cite{special_polynomials} and using the fact that
\begin{equation}
    \int_{-1}^1 dx T_n(x)=\int_0^\pi d\theta\sin\theta \cos(n\theta)=
\end{equation}
\begin{equation*}
    ~~~~~~~~~~~~~~~=\int_0^\pi d\theta\frac{\sin((n+1)\theta)-\sin((n-1)\theta)}{2}=\frac{(-1)^n+1}{1-n^2}(1-\delta_{n,1}) \,,
\end{equation*}
we can solve the following integral:
\begin{equation}\label{proj_int}
    I_{n,l}=\int_0^\pi d\theta \sin\theta \cos(n\theta)P_l(\cos\theta)=
\end{equation}
\begin{equation*}
    ~~~~=\sum\limits_{j=0}^{\floor{l/2}} a_{l,j}\int_{-1}^1 dx T_n(x) T_{l-2j}(x)=
\end{equation*}
\begin{equation*}
    ~~~~=\sum\limits_{j=0}^{\floor{l/2}} \frac{a_{l,j}}{2}\int_{-1}^1 dx \left[T_{n+l-2j}(x) +T_{|n-l+2j|}(x)\right]=
\end{equation*}
\begin{equation*}
    ~~~~=\sum\limits_{j=0}^{\floor{l/2}} a_{l,j}\left[\frac{1}{1-(n+l-2j)^2}+\frac{1}{1-(n-l+2j)^2}\right]\delta_{((n+l)\bmod 2),0}.
\end{equation*}
Because of the orthogonality of Legendre polynomials\cite{special_polynomials},
\begin{equation}
    \int_{-1}^1 dx P_l(x) P_{l'}(x)=\frac{2}{2l+1}\delta_{l,l'}\,,
\end{equation}
the coefficients $b^n_l$ and $a^n_l$ appearing in Eq.~(\ref{cos-sin-legendre0}) can be written in terms of the integrals (\ref{proj_int}) as follows
\begin{equation}
    b^n_l=\left(l+\frac{1}{2}\right)I_{n,l}~,~\qquad a^n_l=-\frac{1}{n}\left(l+\frac{1}{2}\right)I_{n,l}
\end{equation}
Therefore, we get the following alternative expressions for the bra-ket integrals:

\begin{equation}\label{cos_braket_var}
\left\langle l_1, m_1 \right| e^{i k \phi} \cos(n \theta) \left| l_2 , m_2\right\rangle=~~~~~~~~~~~~~~~~~~~~~~~~~~~~~~~~~~~~~~~~~~~~~~~~~~~~~~~~~~~~~~~~
\end{equation}
\begin{equation*}
    =\sum\limits_{l=|l_1-l_2|}^{l_1+l_2} \sum\limits_{j=0}^{\floor{n/2}}\frac{(-1)^{m_1}}{2}\left(n-2j+\frac{1}{2}\right)I_{n,n-2j}\sqrt{\frac{(2 l_1+1)(2 l_2+1)(l+k)!}{(l-k)!}}
\end{equation*}
\begin{equation*}
\times  \left\langle l_1 , -m_1 , l_2 , m_2 \right| \left. l , -k\right\rangle \left\langle l_1 , 0 , l_2 , 0 \right| \left. l , 0\right\rangle I(n-2 j, 0, l,-k)\,,
\end{equation*}
\begin{equation}\label{sin_braket_var}
\left\langle l_1, m_1 \right| e^{i k \phi} \sin(n \theta) \left| l_2 , m_2\right\rangle=~~~~~~~~~~~~~~~~~~~~~~~~~~~~~~~~~~~~~~~~~~~~~~~~~~~~~~~~~~~~~~~~
\end{equation}
\begin{equation*}
=\sum\limits_{l=|l_1-l_2|}^{l_1+l_2} \sum\limits_{j=0}^{\floor{n/2}}\frac{(-1)^{m_1+1}}{2 n}\left(n-2j+\frac{1}{2}\right)I_{n,n-2j}\sqrt{\frac{(2 l_1+1)(2 l_2+1)(l+k)!}{(l-k)!}}
\end{equation*}
\begin{equation*}\times \left\langle l_1 , -m_1 , l_2 , m_2 \right| \left. l , -k\right\rangle \left\langle l_1 , 0 , l_2 , 0 \right| \left. l , 0\right\rangle I(n-2 j, 1, l,-k)\,.
\end{equation*}
This alternative derivation is more straightforward than the one we found, but it gives less compact expressions involving double factorials and an additional sum instead of the gamma functions appearing in our solutions. With these alternative formulas we observe a slightly higher computation time with respect to the expressions we found, which can become relevant when there are many integrals of the type (\ref{bra_ket1}) and (\ref{bra_ket2}) to be computed.

\section{Solutions for the axisymmetric case ($k=0$)}

The formulas appearing in the alternative derivation described in \ref{referee_alternative} can be used also for finding simplified expressions in the axisymmetric case ($k=0$). In this simplified case, the key integrals appearing in Eq.~(\ref{cos_braket0}) involve just Legendre polynomials instead of Legendre associated functions, thus coinciding with integrals \ref{proj_int}. Consequently, Eq.~(\ref{cos_braket0}) for $k=0$ can be reduced to
\begin{equation}\label{cos_braket0k0}
\left\langle l_1, m_1 \right| \cos(n \theta) \left| l_2 , m_2\right\rangle=\sum\limits_{l=|l_1-l_2|}^{l_1+l_2}\sqrt{\left( l_1+\frac{1}{2}\right)\left( l_2+\frac{1}{2}\right)}(-1)^{m_1}
\end{equation}
\begin{equation*}
\times\left\langle l_1 , 0 , l_2 , 0 \right| \left. l , 0\right\rangle\left\langle l_1 , -m_1 , l_2 , m_2 \right| \left. l , 0\right\rangle I_{n,l}\,.
\end{equation*}
We can find a similar procedure also for the sine case, shown in Eq.~(\ref{sin_braket0}). From the trigonometric product-to-sum formulas we have
\begin{equation}\label{product-to-sum}
    \sin\theta\sin(n\theta)=\frac{1}{2}\left[\cos((n-1)\theta)-\cos((n+1)\theta) \right]=
\end{equation}
\begin{equation*}
   ~~~~~~~~~~~~~~~=\frac{T_{n-1}(\cos\theta)-T_{n+1}(\cos\theta)}{2}
\end{equation*}
and, therefore, the key integrals appearing in (\ref{sin_braket0}) become
\begin{equation}
    \int_0^\pi d\theta \sin\theta \sin(n\theta) P_l(\cos\theta)=
\end{equation}
\begin{equation*}
    =\sum\limits_{j=0}^{\floor{l/2}} \frac{a_{l,j}}{2}\int_0^\pi d\theta \left[T_{n-1}(\cos\theta)-T_{n+1}(\cos\theta)\right] T_{l-2j}(x)=
\end{equation*}
\begin{equation*}
    =\sum\limits_{j=0}^{\floor{l/2}}\frac{a_{l,j}}{4}\int_0^\pi d\theta\Bigl[T_{n-1+l-2j}(\cos\theta) -T_{|n-1-l+2j|}(\cos\theta)
\end{equation*}
\begin{equation*}
~~~~~~~~-T_{n+1+l-2j}(\cos\theta)+T_{|n+1-l+2j|}(\cos\theta)\Bigr]=0\,.
\end{equation*}
Hence, equation (\ref{sin_braket0}) in this case reduces to
\begin{equation}
    \left\langle l_1, m_1 \right| \sin(n \theta) \left| l_2 , m_2\right\rangle=0\,,
\end{equation}
which can also be deduced by considering the parity of the functions involved.

\end{appendix}

\section*{Acknowledgments}

G.L. is grateful to Massimo Vaglio and Enrico Cannizzaro for useful discussions on the topic. We acknowledge the financial support provided under the European Union's H2020 ERC, Starting Grant agreement no.~DarkGRA--757480. We also acknowledge support under the MIUR PRIN Grant
2020KR4KN2 "String Theory as a bridge between Gauge
Theories and Quantum Gravity" and FARE programmes (GW- NEXT, CUP: B84I20000100001).

\section*{Bibliography}
\bibliographystyle{utphys}
\bibliography{mybibfile}

\providecommand{\href}[2]{#2}\begingroup\raggedright\begin{thebibliography}{10}

\bibitem{Jackson}
J.~D. Jackson, {\em {Classical electrodynamics; 2nd ed.}}
\newblock Wiley, New York, NY, 1975.
\newblock \url{https://cds.cern.ch/record/100964}.

\bibitem{Sakurai}
J.~J. Sakurai, {\em {Modern quantum mechanics; rev. ed.}}
\newblock Addison-Wesley, Reading, MA, 1994.
\newblock \url{https://cds.cern.ch/record/1167961}.

\bibitem{PoissonWill}
E.~Poisson and C.~M. Will, {\em Gravity: Newtonian, Post-Newtonian,
  Relativistic}.
\newblock Cambridge University Press, 2014.

\bibitem{Pani:2013pma}
P.~Pani, ``{Advanced Methods in Black-Hole Perturbation Theory},''
  \href{http://dx.doi.org/10.1142/S0217751X13400186}{{\em Int. J. Mod. Phys. A}
  {\bfseries 28} (2013) 1340018},
  \href{http://arxiv.org/abs/1305.6759}{{\ttfamily arXiv:1305.6759 [gr-qc]}}.

\bibitem{dolan_time_domain}
S.~R. Dolan, ``Superradiant instabilities of rotating black holes in the time
  domain,'' \href{http://dx.doi.org/10.1103/PhysRevD.87.124026}{{\em Phys. Rev.
  D} {\bfseries 87} (Jun, 2013) 124026}.
  \url{https://link.aps.org/doi/10.1103/PhysRevD.87.124026}.

\bibitem{Baumann_2019}
D.~Baumann, H.~S. Chia, J.~Stout, and L.~ter Haar, ``The spectra of
  gravitational atoms,''
  \href{http://dx.doi.org/10.1088/1475-7516/2019/12/006}{{\em Journal of
  Cosmology and Astroparticle Physics} {\bfseries 2019} no.~12, (Dec, 2019)
  006--006}. \url{https://doi.org/10.1088/1475-7516/2019/12/006}.

\bibitem{Dima_2020}
A.~Dima and E.~Barausse, ``Numerical investigation of plasma-driven
  superradiant instabilities,''
  \href{http://dx.doi.org/10.1088/1361-6382/ab9ce0}{{\em Classical and Quantum
  Gravity} {\bfseries 37} no.~17, (Aug, 2020) 175006}.
  \url{https://doi.org/10.1088/1361-6382/ab9ce0}.

\bibitem{scalar-tensor}
V.~Cardoso, I.~P. Carucci, P.~Pani, and T.~P. Sotiriou, ``Black holes with
  surrounding matter in scalar-tensor theories,''
  \href{http://dx.doi.org/10.1103/PhysRevLett.111.111101}{{\em Phys. Rev.
  Lett.} {\bfseries 111} (Sep, 2013) 111101}.
  \url{https://link.aps.org/doi/10.1103/PhysRevLett.111.111101}.

\bibitem{Cardoso:2013opa}
V.~Cardoso, I.~P. Carucci, P.~Pani, and T.~P. Sotiriou, ``{Matter around Kerr
  black holes in scalar-tensor theories: scalarization and superradiant
  instability},'' \href{http://dx.doi.org/10.1103/PhysRevD.88.044056}{{\em
  Phys. Rev. D} {\bfseries 88} (2013) 044056},
  \href{http://arxiv.org/abs/1305.6936}{{\ttfamily arXiv:1305.6936 [gr-qc]}}.

\bibitem{Lingetti:2022psy}
G.~Lingetti, E.~Cannizzaro, and P.~Pani, ``{Superradiant instabilities by
  accretion disks in scalar-tensor theories},''
  \href{http://dx.doi.org/10.1103/PhysRevD.106.024007}{{\em Phys. Rev. D}
  {\bfseries 106} no.~2, (2022) 024007},
  \href{http://arxiv.org/abs/2204.09335}{{\ttfamily arXiv:2204.09335 [gr-qc]}}.

\bibitem{Brito:2015oca}
R.~Brito, V.~Cardoso, and P.~Pani,
  \href{http://dx.doi.org/10.1007/978-3-319-19000-6}{{\em {Superradiance}: {New
  Frontiers in Black Hole Physics}}}, vol.~971.
\newblock Springer, 2020.
\newblock \href{http://arxiv.org/abs/1501.06570}{{\ttfamily arXiv:1501.06570
  [gr-qc]}}.

\bibitem{Arvanitaki}
A.~Arvanitaki and S.~Dubovsky, ``Exploring the string axiverse with precision
  black hole physics,''
  \href{http://dx.doi.org/10.1103/PhysRevD.83.044026}{{\em Phys. Rev. D}
  {\bfseries 83} (Feb, 2011) 044026}.
  \url{https://link.aps.org/doi/10.1103/PhysRevD.83.044026}.

\bibitem{kodama}
H.~Kodama and H.~Yoshino, ``Axiverse and black hole,''
  \href{http://dx.doi.org/10.1142/S2010194512004199}{{\em International Journal
  of Modern Physics: Conference Series} {\bfseries 07} (2012) 84--115},
  \href{http://arxiv.org/abs/https://doi.org/10.1142/S2010194512004199}{{\ttfamily
  https://doi.org/10.1142/S2010194512004199}}.
  \url{https://doi.org/10.1142/S2010194512004199}.

\bibitem{yoshino}
H.~Yoshino and H.~Kodama, ``{Gravitational radiation from an axion cloud around
  a black hole: Superradiant phase},''
  \href{http://dx.doi.org/10.1093/ptep/ptu029}{{\em Progress of Theoretical and
  Experimental Physics} {\bfseries 2014} no.~4, (04, 2014) },
  \href{http://arxiv.org/abs/https://academic.oup.com/ptep/article-pdf/2014/4/043E02/19300525/ptu029.pdf}{{\ttfamily
  https://academic.oup.com/ptep/article-pdf/2014/4/043E02/19300525/ptu029.pdf}}.
  \url{https://doi.org/10.1093/ptep/ptu029}. 043E02.

\bibitem{BH_bomb}
P.~Pani, V.~Cardoso, L.~Gualtieri, E.~Berti, and A.~Ishibashi, ``Black-hole
  bombs and photon-mass bounds,''
  \href{http://dx.doi.org/10.1103/PhysRevLett.109.131102}{{\em Phys. Rev.
  Lett.} {\bfseries 109} (Sep, 2012) 131102}.
  \url{https://link.aps.org/doi/10.1103/PhysRevLett.109.131102}.

\bibitem{Perturbs_pani}
P.~Pani, V.~Cardoso, L.~Gualtieri, E.~Berti, and A.~Ishibashi, ``Perturbations
  of slowly rotating black holes: Massive vector fields in the kerr metric,''
  \href{http://dx.doi.org/10.1103/PhysRevD.86.104017}{{\em Phys. Rev. D}
  {\bfseries 86} (Nov, 2012) 104017}.
  \url{https://link.aps.org/doi/10.1103/PhysRevD.86.104017}.

\bibitem{Baryakhtar}
M.~Baryakhtar, R.~Lasenby, and M.~Teo, ``Black hole superradiance signatures of
  ultralight vectors,''
  \href{http://dx.doi.org/10.1103/PhysRevD.96.035019}{{\em Phys. Rev. D}
  {\bfseries 96} (Aug, 2017) 035019}.
  \url{https://link.aps.org/doi/10.1103/PhysRevD.96.035019}.

\bibitem{brito_1}
R.~Brito, V.~Cardoso, and P.~Pani, ``Massive spin-2 fields on black hole
  spacetimes: Instability of the schwarzschild and kerr solutions and bounds on
  the graviton mass,'' \href{http://dx.doi.org/10.1103/PhysRevD.88.023514}{{\em
  Phys. Rev. D} {\bfseries 88} (Jul, 2013) 023514}.
  \url{https://link.aps.org/doi/10.1103/PhysRevD.88.023514}.

\bibitem{brito_2}
R.~Brito, S.~Grillo, and P.~Pani, ``Black hole superradiant instability from
  ultralight spin-2 fields,''
  \href{http://dx.doi.org/10.1103/PhysRevLett.124.211101}{{\em Phys. Rev.
  Lett.} {\bfseries 124} (May, 2020) 211101}.
  \url{https://link.aps.org/doi/10.1103/PhysRevLett.124.211101}.

\bibitem{frolov_proca}
V.~P. Frolov, P.~Krtou\ifmmode~\check{s}\else \v{s}\fi{},
  D.~Kubiz\ifmmode~\check{n}\else \v{n}\fi{}\'ak, and J.~E. Santos, ``Massive
  vector fields in rotating black-hole spacetimes: Separability and quasinormal
  modes,'' \href{http://dx.doi.org/10.1103/PhysRevLett.120.231103}{{\em Phys.
  Rev. Lett.} {\bfseries 120} (Jun, 2018) 231103}.
  \url{https://link.aps.org/doi/10.1103/PhysRevLett.120.231103}.

\bibitem{Dolan:2012yt}
S.~R. Dolan, ``{Superradiant instabilities of rotating black holes in the time
  domain},'' \href{http://dx.doi.org/10.1103/PhysRevD.87.124026}{{\em
  Phys.Rev.} {\bfseries D87} (2013) 124026},
\href{http://arxiv.org/abs/1212.1477}{{\ttfamily arXiv:1212.1477 [gr-qc]}}.

\bibitem{LingettiThesis}
G.~Lingetti, ``{PhD thesis, to be submitted (2023)},''.

\bibitem{Fujii:2003pa}
Y.~Fujii and K.~Maeda, \href{http://dx.doi.org/10.1017/CBO9780511535093}{{\em
  {The scalar-tensor theory of gravitation}}}.
\newblock Cambridge Monographs on Mathematical Physics. Cambridge University
  Press, 7, 2007.

\bibitem{Dima:2020rzg}
A.~Dima and E.~Barausse, ``{Numerical investigation of plasma-driven
  superradiant instabilities},''
  \href{http://dx.doi.org/10.1088/1361-6382/ab9ce0}{{\em Class. Quant. Grav.}
  {\bfseries 37} no.~17, (2020) 175006},
  \href{http://arxiv.org/abs/2001.11484}{{\ttfamily arXiv:2001.11484 [gr-qc]}}.

\bibitem{Cardoso:2013fwa}
V.~Cardoso, I.~P. Carucci, P.~Pani, and T.~P. Sotiriou, ``{Black holes with
  surrounding matter in scalar-tensor theories},''
  \href{http://dx.doi.org/10.1103/PhysRevLett.111.111101}{{\em Phys. Rev.
  Lett.} {\bfseries 111} (2013) 111101},
  \href{http://arxiv.org/abs/1308.6587}{{\ttfamily arXiv:1308.6587 [gr-qc]}}.

\bibitem{rose}
M.~Rose, {\em Elementary Theory of Angular Momentum}.
\newblock New York, 1957.
\newblock \url{https://books.google.it/books?id=MyEFzQEACAAJ}.

\bibitem{math_meth}
{\em Series Expansions of Arbitrary Functions},
  \href{http://dx.doi.org/https://doi.org/10.1002/9783527617210.ch2}{ch.~2,
  pp.~48--111}.
\newblock John Wiley and Sons, Ltd, 1989.
\newblock
  \url{https://onlinelibrary.wiley.com/doi/abs/10.1002/9783527617210.ch2}.

\bibitem{special_polynomials}
M.~E.~H. Ismail, \href{http://dx.doi.org/10.1017/CBO9781107325982}{{\em
  Classical and Quantum Orthogonal Polynomials in One Variable}}.
\newblock Encyclopedia of Mathematics and its Applications. Cambridge
  University Press, 2005.

\bibitem{legendre_integral}
H.~Mavromatis and R.~Alassar, ``A generalized formula for the integral of three
  associated legendre polynomials,''
  \href{http://dx.doi.org/https://doi.org/10.1016/S0893-9659(98)00180-3}{{\em
  Applied Mathematics Letters} {\bfseries 12} no.~3, (1999) 101--105}.
  \url{https://www.sciencedirect.com/science/article/pii/S0893965998001803}.

\bibitem{whittaker_watson_1996}
E.~T. Whittaker and G.~N. Watson,
  \href{http://dx.doi.org/10.1017/CBO9780511608759}{{\em A Course of Modern
  Analysis}}.
\newblock Cambridge Mathematical Library. Cambridge University Press, 4~ed.,
  1996.

\end{thebibliography}\endgroup

\end{document}